\documentclass[preprint]{aastex62}

\newcommand{\kms}{\,km\,s$^{-1}$}
\newcommand{\Msol}{$M_\odot$}

\newcommand{\Hethree}{$^3$He}

\newcommand{\Mv}{$M_{\rm V}$}
\newcommand{\Ha}{H${\alpha}$}

\newcommand{\Hg}{H${\gamma}$}

\newcommand{\vrad}{$v_{\rm rad}$}

\newcommand{\BeII}{$^7$\ion{Be}{2}}
\newcommand{\Be}{$^7$Be}
\newcommand{\LiI}{\ion{Li}{1}}
\newcommand{\Li}{$^7$Li}
\newcommand{\FeII}{\ion{Fe}{2}}
\newcommand{\CaII}{\ion{Ca}{2}}
\newcommand{\CaK}{\ion{Ca}{2} K}

\newcommand{\NLiH}{$N({\rm ^{7}Li})/N({\rm H})$} 
\newcommand{\NLiHfinal}{$N({\rm ^{7}Li})/N({\rm H})_{\rm final}$} 
\newcommand{\NBeH}{$N({\rm ^{7}Be})/N({\rm H})$} 
\newcommand{\NBeCa}{$N({\rm ^{7}Be})/N({\rm Ca})$} 
\newcommand{\NCaH}{$N({\rm Ca})/N({\rm H})$} 
\newcommand{\XLiH}{$X({\rm ^{7}Li})/X({\rm H})$}

\newcommand{\IRAF}{the Image Reduction and Analysis Facility (IRAF)}

\newcommand{\websmarts}{http://www.astro.sunysb.edu/fwalter/SMARTS/NovaAtlas/atlas.html}
\newcommand{\iraf}{IRAF is distributed by the National Optical Astronomy Observatory, which is operated by the Association of Universities for Research in Astronomy (AURA) under a cooperative agreement with the National Science Foundation.}
\newcommand{\hdsred}{https://www.naoj.org/Observing/Instruments/HDS/hdsql-e.html}

\graphicspath{{./}{figures/}}

\shorttitle{Detection of \BeII\ in \object{V5669 Sgr}}
\shortauthors{Arai et al.}

\begin{document}

\title{Detection of \BeII\ in the Classical Nova \object{V5669 Sgr} (Nova Sagittarii 2015 No.3)}

\correspondingauthor{Akira Arai}
\email{arai6a@cc.kyoto-su.ac.jp}
\author[0000-0002-5756-067X]{Akira Arai}
\affil{Koyama Astronomical Observatory, Kyoto Sangyo University, Motoyama, Kamigamo, Kita-ku, Kyoto-city, Kyoto, Japan, 803-8555, Japan}
\author[0000-0001-8813-9338]{Akito Tajitsu}
\affil{Subaru Telescope, National Astronomical Observatory of Japan, 650 North A’ohoku Place, Hilo, HI 96720, USA}
\author[0000-0003-2011-9159]{Hideyo Kawakita}
\affil{Koyama Astronomical Observatory, Kyoto Sangyo University, Motoyama, Kamigamo, Kita-ku, Kyoto-city, Kyoto, Japan, 803-8555, Japan}
\author[0000-0003-4490-9307]{Yoshiharu Shinnaka}
\affil{Koyama Astronomical Observatory, Kyoto Sangyo University, Motoyama, Kamigamo, Kita-ku, Kyoto-city, Kyoto, Japan, 803-8555, Japan}

\begin{abstract}
We report the new detection of \BeII\ in the ultraviolet spectra of \object{V5669 Sgr} during its early decline phase ($+24$ and $+28$ d).  
We identified three blue-shifted absorption systems in our spectra.  
The first two, referred to as low- and high-velocity components, were noticeably identified among \ion{H}{1} Balmer, \ion{Na}{1}~D, and \ion{Fe}{2} whose lower energies of transients are low ($<4$ eV). 
The third absorption component was identified among \ion{N}{2}, \ion{He}{1}, and \ion{C}{2} lines whose lower energy levels are relatively high ($9$--$21$ eV).  
The absorption lines of \BeII\ at $3130.583$ \AA, and $3132.228$ \AA\ were identified as the first and second components in our observations.  
No evidence suggested the existence of \LiI\ at $6708$\,\AA\ in any velocity components. 
The estimated number density ratio of lithium relative to hydrogen, which was finally produced by this object using the equivalent widths of \Be\ and \ion{Ca}{2}~K, \NLiHfinal\,, is $4.0\pm0.7\times10^{-6}$.  
This value is an order of magnitude lower than the average observed values for classical novae wherein \Be\ has been detected, 
and is comparable to the most optimistic value of theoretical predictions. 
\end{abstract}


\section{Introduction \label{introduction}}

The \Li\ production in classical novae was first confirmed in \object{V339 Del} by \citet{tajitsu15}. 
They reported the detection of blue-shifted absorption lines of the resonance lines of \BeII\ at $3130.583$ \AA\ and $3132.228$ \AA\ in near-ultraviolet (UV) high-resolution spectra. 
\Be\ is an unstable isotope that decays to form \Li\ through the electron capture within a half-life of $53.22$ days.
Therefore, the detection of \BeII\ lines in the nova spectra provides crucial evidence of \Li\ production by nova explosion.
Since the first confirmation of \Li\ production, \BeII\ and/or \LiI\ $6708$ \AA\ have been directly detected in several classical novae through high-resolution spectroscopic observations.
The first detection of \LiI\ at $6708$ \AA\ was reported in \object{V1369 Cen} \citep{izzo15}.
The second and third detections of \BeII\ were reported in \object{V5669 Sgr} \citep{molaro16,tajitsu16} and \object{V2944 Oph} \citep{tajitsu16}, respectively. 
These three classical novae showed slow declines in their light curves, which suggests that they are CO novae.
Furthermore, \citet{izzo18} reported \BeII\ detection in a very fast nova \object{V407 Lup}. 
For another very fast nova, \object{V838 Her}, \cite{selvelli18} reported the existence of absorption and emission features of \BeII\ in the historical UV spectra obtained by {\it the International Ultraviolet Explorer} (\citealt{boggess78}). 
The very rapid decline in their light curves (duration up to $+2$ mag after the maximum, $t_{2} \sim 5$ and $2$ d for \object{V407 Lup} and \object{V838 Her}, respectively) and characteristics of their spectra indicate that they are ONe novae.
Recently \cite{molaro2020} reported further detections of \BeII\ and/or \LiI\ in \object{Nova Mus 2018} and \object{ASASSN-18fv}. 
These two classical novae are probably CO novae because their light curves indicate that they are moderately fast or slow novae. 
\cite{molaro2020} summarized that the abundance (number) ratio of \Li\ to hydrogen enhanced by the above-observed novae, \NLiH\,, is in the range of $1.5$--$17 \times 10^{-5}$ (average of  $5.4 \times 10^{-5}$). 
No major difference exists in the \NLiH\ ratio among CO and ONe novae.
This observational fact implies that \Li\ production widely occurs in classical novae. 
Using recent observational results, \citet{cescutti19} demonstrated that Li production in classical novae could mostly explain the Galactic Li evolution curve without remarkable contributions from other source candidates (e.g., red giants and AGB stars).

In studies of \Li\ production in classical novae,  theoretical predictions preceded observations. 
\citet{cameron71} first established \Li\ production in red giants via the 
\Hethree($\alpha$,$\nu$)\Be\ reaction,  and since then, a number of efforts have been undertaken on thermonuclear runaway (TNR) in classical novae.
\cite{arnould75} calculated the \Li\ production rate for several combinations of fixed temperatures and densities to study the resulting nucelosynthesis. 
\cite{starrfield78} used their hydrodynamic code to investigate nucleosynthesis during a TNR in a solar-abundance envelope on a white   dwarf (WD) and found that \Li\ can be overproduced. 
\citet{jose98} conducted realistic hydrodynamic calculations that took into account the temperature development in the envelope during the accretion phase before the onset of the TNR and predicted the Li production during the TNR, obtaining values of \NLiH\ of $\lesssim 4.7 \times 10^{-6}$ and $\lesssim 4.8 \times 10^{-6}$ for ONe novae and CO novae, respectively.
Several studies on Li evolution in the galaxy that draw on the results of such numerical simulations have concluded that classical novae may contribute up to $10$\% of the Li produced in the recent universe \citep{hernanz96,prantzos12}.
Recently, using the Modules for Experiments in Stellar Astrophysics (MESA) code (e.g., \citealt{paxton11}), \cite{rukeya17} reported the prediction of \Li\ production similar to those of \cite{jose98}.
\cite{starrfield20} obtained the largest \Li\ production obtained to date for a $1.35$\, \Msol CO WD [\NBeH\ $= 7.0 \times 10^{-6}$], assuming that mixing between the core and solar-abundance material in the envelope gas transferred from the companion star switches  on immediately after the onset of the TNR.
However, such estimates of \NBeH\ still amount to only several tenths of derived from the observations.
\cite{jose20} developed a more realistic model, which incorporates mixing and convection extracted directly from their own 3D simulations back into a 1D model.
Their results suggest that the \Li\ production is much less than that obtained in the previous works (e.g., \citealt{rukeya17} and \citealt{starrfield20}); however, the abundances they obtained for the intermediate-mass elements agree with the observations.
In addition to the large disagreement between the observations and the theoretical predictions, large variations exist among the theoretical predictions of the amounts of \Li\ produced. 
This suggests that the results may depend strongly on differences in the handling of mixing and convection and of the accretion phase before the TNR.

The recent observational facts strongly suggest that classical novae are the major sources of Li in the galaxy.
However, some questions on this phenomenon still remain. 
The first question concerns the discrepancy between the observed abundances and theoretical predictions as described above.
The appearance rate of Li production is another question.
Recently, \citet{molaro2020} reported the detection of neither \BeII\ nor \LiI\ in the post-outburst spectra of \object{ASASSN-17hx} and \object{Nova Cir 2018}.
This is the first nondetection report of \BeII\ or \LiI\ in classical novae after \citet{tajitsu15}, and it suggests that there could be a large scattering of \Li\ production among classical novae. 
Solving these questions is essential for more precise understandings of the contributions of classical novae to Galactic Li evolution.  
Therefore, further observations of \BeII\ (and \LiI) in classical novae during their early decline phases are required. 
In this study, we report a new detection of \BeII\ in \object{V5669 Sgr} (\object{Nova Sgr 2015 No.3}).

\section{V5669 Sgr (Nova Sagitarii 2015 No.3)\label{v5669sgr}}


\begin{deluxetable}{lll}[b]
\tablecaption{Information of V5669~Sgr.\label{tab:target}}
\tablecolumns{3}
\tablewidth{0pt}
\tablehead{
\colhead{Item} & \colhead{Value} }
\startdata
Discovery date & UT 2015 September 27.4293\\
               & MJD 57292.4293\\
Coordinate ${(\rm{J2000.0})}$ & $\alpha = 18^{\rm h}$03$^{\rm m}$32$^{\rm s}$.75, $\delta=-28^{\rm o}16{\rm '}05{\rm ''}.4$\\
$t_{\max}$        & 2015 September 29.1 (MJD 57294.1)    \\
$V_{\max}$        & $8.572 \pm 0.001$ mag\\
$(B-V)_{\rm max}$ & $0.793 \pm 0.001$ mag\\ 
$t_2$             & $24$ days      \\
$E_{B-V}$$^{*}$  & $0.44\pm0.02$ mag\\
\Mv$^{*}$         & $-7.8\pm0.4$ mag\\
$d^{*}$           & $10.0\pm1.9$ kpc\\
\enddata
\tablecomments{$\ast$: The estimation of the interstellar extinction ($E_{B-V}$), the absolute magnitudes (\Mv), and the distance ($d$) are shown in Appendix~\ref{appendix}.}
\end{deluxetable}

\object{V5669 Sgr} (Nova Sgr 2015 No.3) was discovered as a possible transient object, \object{PNV J18033275-2816054}, by Koichi Itagaki on UT $27.429$ September 2015 at $9.8$ mag (unfiltered) using a $0.5$ m Schmidt Cassegrain telescope with an unfiltered CCD camera at his private observatory \citep{nakano15}. 
The earliest spectroscopic observation by \cite{fujii15} with a low-resolution spectrograph ($R \sim 500$) on a $40$ cm telescope at the Fujii-Kurosaki Observatory (in Okayama, Japan) was performed around $1.4$ h after the discovery, and his spectra show \ion{H}{1} Balmer, \FeII\ and \ion{Na}{1} emission lines with P-Cygni profiles, suggesting that the expanding velocity was approximately $1100$\,\kms.  
\cite{williams15}  also reported spectra taken by the FRODOSpec \citep{barnsley12} on the $2.0$ m Liverpool Telescope (in Spain) around $10$ h after the discovery, which showed remarkable lines of \ion{H}{1}, \FeII, \ion{O}{1} $\lambda7773$, $\lambda8446$, and \ion{He}{1}\,$\lambda6678$.  
The emission and absorption features of \Ha\ indicate that the expansion velocity was around $2000$\,\kms.
They concluded that the spectral type \citep{williams92} of the nova was classified into \FeII\ class.

\begin{figure}[th] 
\epsscale{0.8}
\plotone{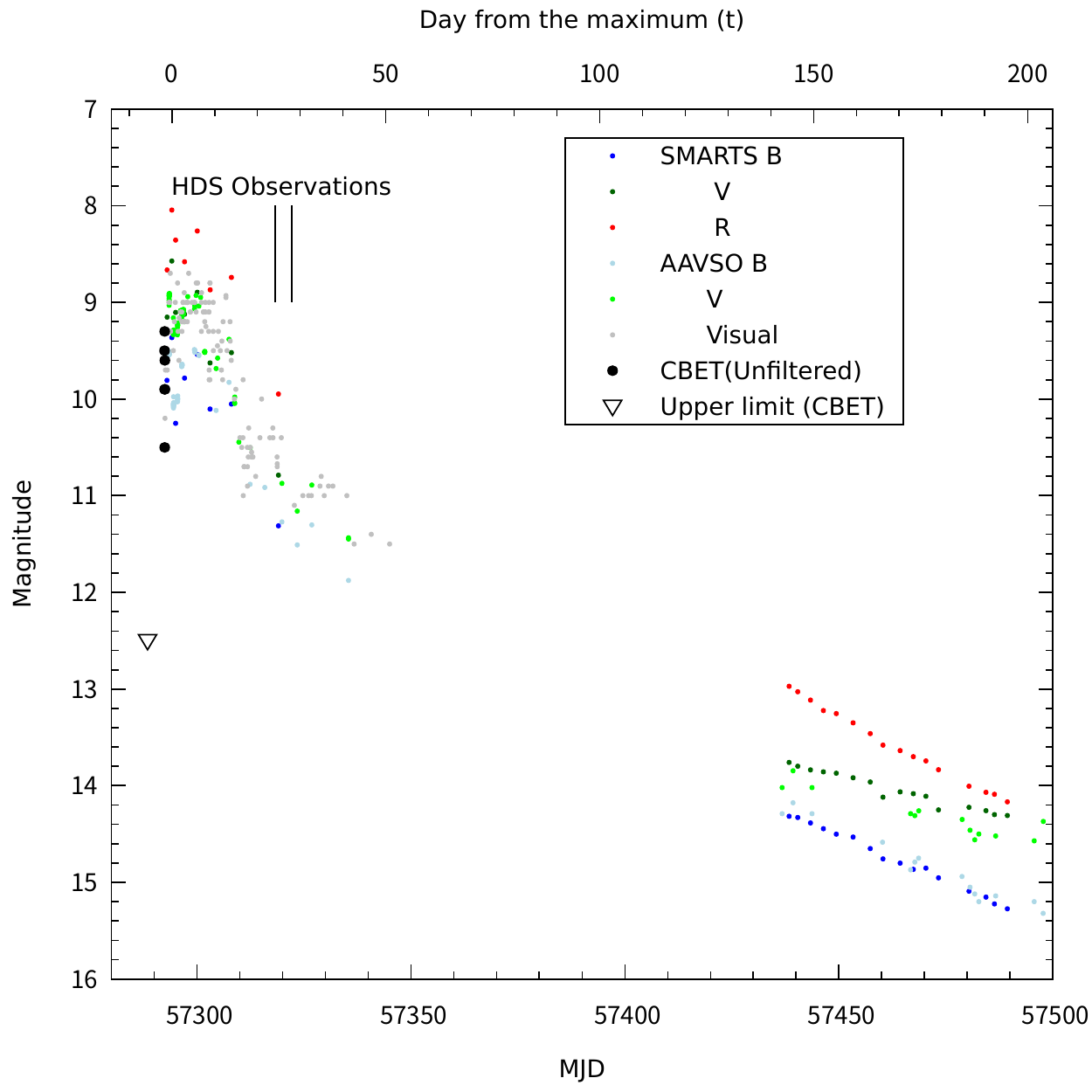}
\caption{Light curves of \object{V5669 Sgr}. Photometric data were quoted from \citet[CBET]{nakano15}, the AAVSO Archives, and the Stony Brook/SMARTS Atlas. Vertical lines denote the time of our observations on $+24$ and $+28$\,d. \label{fig:lightcurve}}
\end{figure}

Figure~\ref{fig:lightcurve} displays the optical light curves of \object{V5669 Sgr}\footnote{Photometric data are taken from \citet{nakano15} the AAVSO Archives (Kafka, S., 2019, Observations from the AAVSO International Database,  https://www.aavso.org),and The Stony Brook/SMARTS Atlas of Southern Novae \citep[\websmarts][]{walter12}.}.
The object was discovered during the early rising phase about $1.6$\,d before the maximum.  
We set the first maximum date to MJD $57294.1$ ($= t_0$) at $V=8.57$ mag based on data in the Stony Brook/SMARTS Atlas.  
The basic parameters of the nova are listed in Table~\ref{tab:target}.   
The estimations of the extinction, $E_{B-V}$, the absolute magnitude, $M_{V}$, and the distance, $d$, are shown in Appendix~\ref{appendix}  using the SMARTS data set and measurements of diffuse interstellar bands detected in our high-resolution spectra.

\object{V5669 Sgr} is potentially a CO nova as it shows a moderately fast decline in its brightness with $t_2 = 24$\,d. 
Strong \ion{Fe}{2}, \ion{Na}{1}, or \ion{O}{1} lines associated with the P-Cygni profile are dominant in the initial spectra described above and also in our spectra (see Section~\ref{results}).  
Note that He/N-type emission features, which are often observed in early decline phases of ONe novae, are not prominent among them.

After its explosion, \object{V5669 Sgr} was detected in the GAIA DR2 catalog as a source of $G=14.3$ (GAIA DR2 4062508582302805760) $0.16$ arcsec apart from the position of the nova. 
Its parallax and distance are not available in the catalog.  
Moreover, no counterparts are present in the Pan-STARRS catalog \citep{PanSTARRS2016}.   
The nearest red star in the Pan-STARRS catalog was found at $0.9$ arcsec apart from the nova (PSO J180332.784-281604.706, $g=19.02\pm0.04$, $r=18.102\pm0.003$), 
whose counterpart can be found in the GAIA DR2 catalog (GAIA DR2 4062508582302826240, $G=18.25$).  
Thus, we concluded that the progenitor is not detected in the Pan-STARRS and previous surveys.


\begin{deluxetable}{llccccc}
\tablecaption{Journal of Observations of \object{V5669~Sgr} by Subaru/HDS \label{tab:obs}}
\tablecolumns{7}
\tablewidth{0pt}
\tablehead{
\colhead{Date} & \colhead{MJD} & \colhead{t}     & \colhead{Seup} & \colhead{Exp. time} &\colhead{Airmass}\\
\colhead{}     & \colhead{}    & \colhead{(day)} & \colhead{}     & \colhead{(sec)}         &\colhead{}
}
\startdata
23 Oct 2015 & $57318.19$ & $+24.09$  & Ub & $3000$ & $1.88$\\
23 Oct 2015 & $57318.23$ & $+24.13$  & Yb & $1800$ & $2.58$\\
27 Oct 2015 & $57322.19$ & $+28.09$  & Ub & $2400$ & $1.97$\\
27 Oct 2015 & $57322.22$ & $+28.12$  & Yb & $1800$ & $2.59$\\
\enddata
\end{deluxetable}

\section{Observations and Reductions \label{observations}}

We performed optical high dispersion spectroscopic observations for \object{V5669 Sgr} using the High Dispersion Spectrograph \citep[HDS;][]{noguchi02} mounted on the 8.2 m Subaru telescope and succeeded at two epochs ($+24$ and $+28$\,d) during its rapid decline phase.  
Table \ref{tab:obs} shows a summary of our HDS observations. 
Our spectra cover $3020$--$6865$\,\AA\ with two different grating settings of the spectrograph:  Setup-Ub, $3020$--$4631$\,\AA\ with a gap at $3748$--$3790$\,\AA, and Setup-Yb, $4108$--$6865$\,\AA\ with a gap at $5434$--$5514$\,\AA. 
The spectral resolving power was $\sim45,000$ with a $0.80$'' ($0.4$\,mm) slit width. 
Data reduction was conducted using  \IRAF\footnote{\iraf} software with the standard manner for HDS data\footnote{\hdsred}.  
The non-linearity of the CCD pixels was corrected using the method described by \cite{tajitsu10}. 
The wavelength calibration was performed using the Th-Ar comparison spectra.   
The typical residual in the wavelength calibration was approximately $10^{-3}$\,\AA\ ($\sim 0.1$\,\kms).   
The observed spectra were converted to the heliocentric wavelength scale.  
Spectro-photometric calibrations were performed with the spectra of \object{$\sigma$ Sgr} ($V = 2.058$, B2V \footnote{These values are taken from Simbad \citep{simbad}.}) obtained nearly at the same airmass of the nova on the same nights.
The telluric absorption correction was not performed.

In principle, we refer to \cite{NIST_ASD} for wavelengths, log($gf$)s, and lower and upper transition energies ($E_{\rm low}$ and $E_{\rm up}$) of each ion herein. 
Multiplet numbers are taken from \cite{moore43}. 
We refer to such parameters of \ion{Cr}{2}~(5) lines from \cite{nilsson06}.

\section{Results\label{results}}

\subsection{Overview \label{overview}\label{sect4.1}}

\begin{figure*} 
\epsscale{1.0}
\plotone{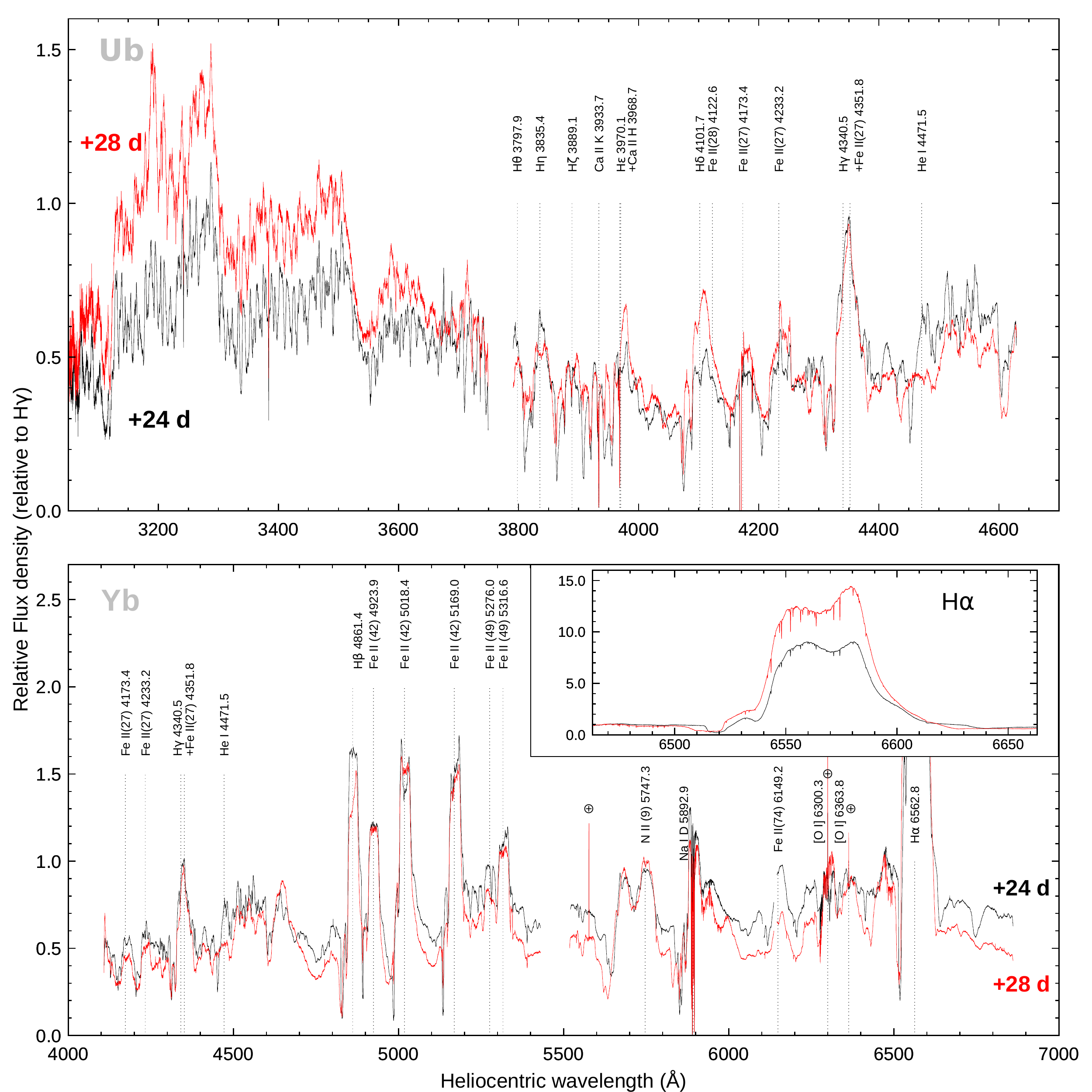}
\caption{Overall spectra of \object{V5669 Sgr} on $+24$\,d (black) and $+28$\,d (red).  The upper and lower panel show ``Ub'' region spectra ($3050$--$4631$\,\AA), and ``Yb''-setting spectra ($4108$--$6865$\,\AA), respectively.  The vertical axis denotes the relative flux density normalized with the peak of emission components of \Hg\ included in both regions.  Dotted lines show the line identifications of strong emission lines.  Crossed circles indicate the telluric emission lines. The magnified view in \Ha\ is shown in the inset in the lower panel.\label{fig:overview}}
\end{figure*}

Figure~\ref{fig:overview} shows the overall spectra of \object{V5669 Sgr} obtained on $+24$  and $+28$\,d. 
The identifications of strong emission lines are listed in Table~\ref{tab:emission_linelist}.  
The emission lines originating from the \ion{H}{1} Balmer series, \FeII, and \CaII\ are easily identified in the spectra.
These strong emission lines are accompanied by P-Cygni-like profiles, and the mean FWHMs of these emission lines are approximately $2000$\,\kms.  
Each \ion{H}{1} emission line has a saddle-shaped profile, as seen in the \Ha\ emission line shown in the inset in the panel of Figure~\ref{fig:overview}.  
These spectral features indicate that the nova is an ordinary \FeII-type classical nova with a CO WD as reported in previous studies \citep{fujii15,williams15}.

\begin{deluxetable}{llc|llc}
\tablecaption{List of Identified Emission Lines in Figure~\ref{fig:overview}. $\lambda_{\rm obs}$ is the observed wavelength around the center of each emission line on $+24$\,d. $\lambda_{\rm rest}$ is the laboratory wavelength.  The multiplet numbers are taken from \cite{moore43}\label{tab:emission_linelist}.  The left curly brackets denote the observed blended lines in the column between $\lambda_{\rm rest}$ and $\lambda_{\rm lab}$.}
\tablecolumns{8}
\tablewidth{0pt}
\tablehead{
\colhead{$\lambda_{\rm obs}$(\AA)} & \colhead{$\lambda_{\rm lab}$(\AA)} & \colhead{Line} &  \colhead{$\lambda_{\rm obs}$(\AA)} & \colhead{$\lambda_{\rm lab}$(\AA)} & \colhead{Line}  
}
\startdata
3794.2                  & 3797.909   & H{$\theta$}            & 4864.6 & 4861.350 & H{$\beta$}       \\
3835.3                  & 3835.397   & H{$\eta$}              & 4926.0 & 4923.921 & \ion{Fe}{2} (42) \\     
3888.1                  & 3889.064   & H{$\zeta$}             & 5022.0 & 5018.436 & \ion{Fe}{2} (42) \\     
3930.4                  & 3933.68    & \ion{Ca}{2} K          & 5173.0 & 5169.028 & \ion{Fe}{2} (42) \\    
3972.5                  & 3968.66    & \ion{Ca}{2} H          & 5278.2 & 5275.997 & \ion{Fe}{2} (49) \\   
                        & +3970.075 & H{$\epsilon$}           & 5320.7 & 5316.609 & \ion{Fe}{2} (49) \\         
4102.5                  & 4101.734   & H{$\delta$}            & 5749.1 & 5747.300 & \ion{N}{2} (9)   \\   
4125.1                  & 4122.638   & \ion{Fe}{2} (28)       & 5894.3 & 5892.937 & \ion{Na}{1} D    \\     
4177.6                  & 4173.450   & \ion{Fe}{2} (27)       & 6152.1 & 6149.246 & \ion{Fe}{2} (74) \\ 
4239.5                  & 4233.160   & \ion{Fe}{2} (27)       & 6303.0 & 6300.304 & [O I]            \\ 
4350.6                  & 4340.472   & H{$\gamma$}            & 6361.3 & 6363.776 & [O I] \\   
                        & +4351.763 & \ion{Fe}{2} (27)        & 6567.9 & 6562.79  & H{$\alpha$}      \\   
4473.4                  & 4471.48    & \ion{He}{1}            &        &          &                   \\         
\enddata
\end{deluxetable}

\subsection{Blueshifted Absorption Line Systems (Low- and High-Velocity Components)\label{sect4.2}}
Figure~\ref{fig3} displays the enlarged views of the spectra taken on $+24$\,d (left) and $+28$\,d (right) in the vicinities of H$\gamma$, \ion{Ca}{2} K, \ion{Fe}{2} (42) $\lambda5169.03$, \ion{Na}{1} D$_2$, and \BeII\ $\lambda3131.228$.  
Two blue-shifted absorption components accompany each of the first three lines in both epochs. 
Hereafter, we refer to them as the low-velocity component (LVC) and high-velocity component (HVC).
On $+24$\,d, the LVC (\vrad\ $= -840$ to $-1260$\,\kms) and HVC (\vrad\ $= -1800$ to $-2200$\,\kms) are very distinct. 
Until $+28$\,d, the HVC definitely accelerated to \vrad\ $= -1850$ to $-2500$\,\kms. 
The LVC had a somewhat mild acceleration to \vrad\ $= -840$ to $-1300$\,\kms.
Moreover, LVC and HVC of \ion{Na}{1} D, and \BeII\ can be identified in both epochs, although some contaminations arising from other metal lines are expected.
The strengths of most blue-shifted absorption lines became weaker from $+24$ to $+28$\,d. 
The decrease in the equivalent width ($W$), $\Delta W = -0.88\pm0.07$\,\AA\ for the LVC, and $-2.73\pm0.06$\,\AA\ for the HVC, is  observed clearly for \ion{Fe}{2} $\lambda5169.028$ ($E_{\rm low} = 2.89$ eV). 
The similar blueshifted absorption line systems are occasionally observed in previous novae, for example, \object{V2659 Cyg} \citep{arai16}.  
As in \object{V2659 Cyg}, numerous absorption lines can be identified with LVC originating from singly ionized Fe-peak elements  (\ion{Ti}{2}, \ion{Cr}{2}, and \ion{Mn}{2}).
Most of them do not accompany their emission lines.
They are essentially the same as the ``transient heavy element absorption (THEA)'', proposed and discussed in \cite{williams08}.

\subsection{Another Blueshifted Absorption System (Third Velocity Component)\label{sect4.3}}
Figure~\ref{fig4} shows representative spectra in the vicinities of \ion{N}{2}\,(3) $\lambda5666.63$ and \ion{He}{1} $\lambda5875.621, \lambda6678.151$ on $+24$ and $+28$\,d. 
Absorption lines accompanied by these lines are identified as a single component.
On $+24$\,d, the velocity range of this component, $v_{\rm rad} = -1450$ to $-2070$\,\kms, is similar to that of HVC in this epoch.
On $+28$\,d, however, the component accelerated up to $-2100$ to $-2650$ \kms, which is different from the velocities of both LVC and HVC in this epoch.
The absorption lines of this component have somewhat smooth, round-shaped profiles compared with those in LVC and HVC displayed in Figure~\ref{fig3}. 
This absorption system is not accompanied by any emission lines except for \ion{N}{2} (3) $\lambda5666.63$ ($E_{\rm low}=18.47$\,eV) and \ion{He}{1} $\lambda5875.621$ blending with \ion{Na}{1} D.
Several \ion{N}{2} lines (e.g., $\lambda5747.30$) have absorption lines corresponding to this system. 
The $E_{\rm low}$ of these lines is in the range of $9.3$--$21$\,eV, which is significantly higher than that in the LVC and HVC systems reported in Section~\ref{sect4.2}. 
Considering these characteristics, we conclude that this blue-shifted absorption component is the third component that is neither LVC nor HVC. 
Hereafter we name it the third velocity component (TVC) for convenience. 
We note that the TVC in \object{V5669 Sgr} is similar to the absorption components of \ion{N}{2} and \ion{He}{1} identified in \object{V407 Lup} \citep{izzo18}. 
In \object{V407 Lup}, the radial velocity of this velocity component reached $-3830$\,\kms\ in contrast to the low velocities of \ion{Fe}{2} and \ion{H}{1} lines ($\sim-2300$\,\kms).

\begin{deluxetable}{ccclc}
\tablecaption{Candidates of Lines Contaminating \BeII.\label{tab:contami}}
\tablecolumns{6}
\tablewidth{0pt}
\tablehead{
\colhead{Line} & \colhead{Rest wavelength(\AA)} & \colhead{log($gf$)$^{*}$} & \colhead{$W$(\AA)$^{\dagger}$} & \colhead{Contamination}}
\startdata
\ion{Cr}{2}\,(5) & $3118.649$ & $-0.102$ & {\it 0.24}     & HVC\\
\ion{Cr}{2}\,(5) & $3120.369$ & $+0.108$ & {\it 0.38}     & HVC\\
\ion{Cr}{2}\,(5) & $3124.978$ & $+0.286$ & 0.55$\pm$0.02  & \\
\ion{Cr}{2}\,(5) & $3128.700$ & $-0.528$ & {\it 0.09}     & LVC\\
\ion{Cr}{2}\,(5) & $3132.056$ & $+0.437$ & {\it 0.82}     & LVC\\
\enddata
\tablecomments{
$\ast$: All log($gf$) values of \ion{Cr}{2} (5) lines are quoted from \cite{nilsson06}. \\
${\dagger}$: Measured equivalent widths is only for \ion{Cr}{2}$\lambda3124.978$. $W$s for other lines (italic letters) are calculated  using their log($gf$) values.\\
We regarded that  \ion{Cr}{2}\,(5)\,$\lambda3128.700$ contaminates \BeII\ with half of its $W$ [ = $0.5 \times W_{\rm \lambda} $(\ion{Cr}{2} $\lambda3128.700$)].}
\end{deluxetable}  

\subsection{\BeII\ Abundance\label{sect4.4}}

\begin{figure*} 
\epsscale{1.15}
\plottwo{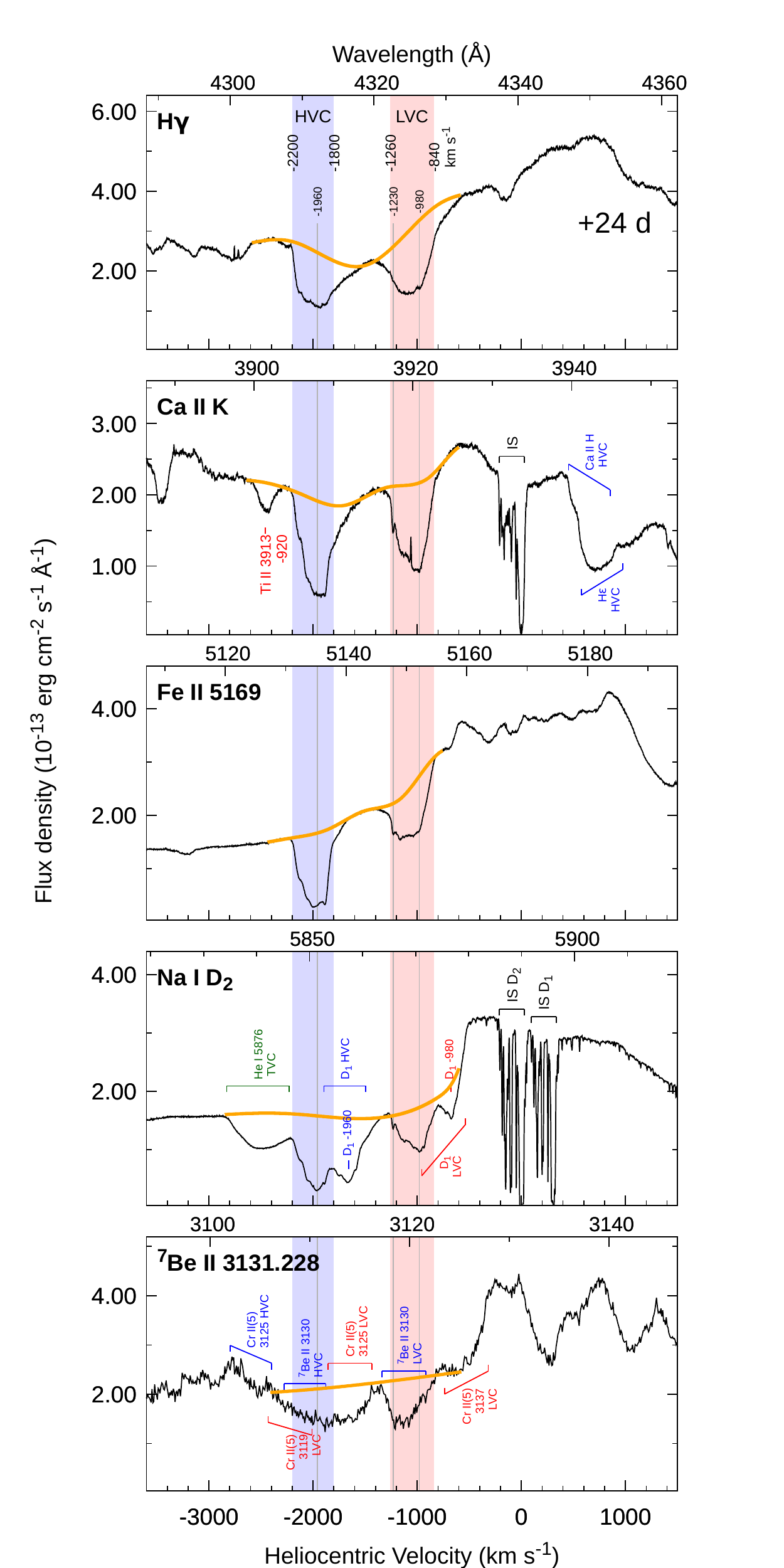}{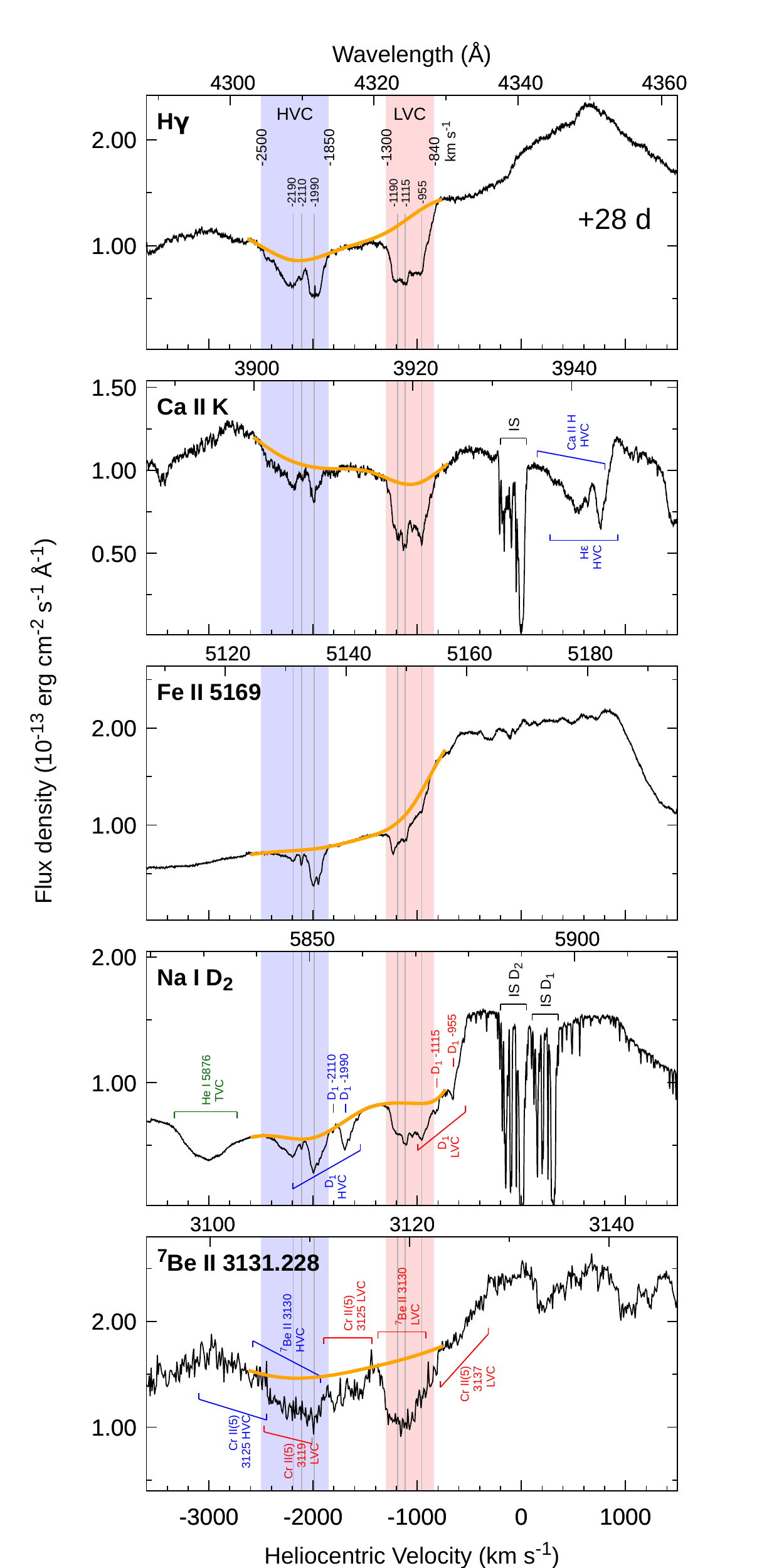}
\caption{Velocity components in the major lines (\Hg, \ion{Ca}{2} K, \BeII\, \ion{Na}{1} D${_2}$, and \ion{Fe}{2} $\lambda5169.028$) on $+24$\,d (left) and $+28$\,d (right).   Orange curves indicate continuum fitting curves for these lines.  Red and blue shaded regions indicate ranges of the LVC and HVC, respectively. Gray vertical lines are subcomponents in these absorption systems determined from \Hg, \ion{Fe}{2} $\lambda$5196.028, and \ion{Ca}{2} K. \label{fig3}}
\end{figure*}

\begin{figure} 
\epsscale{1.15}
\plottwo{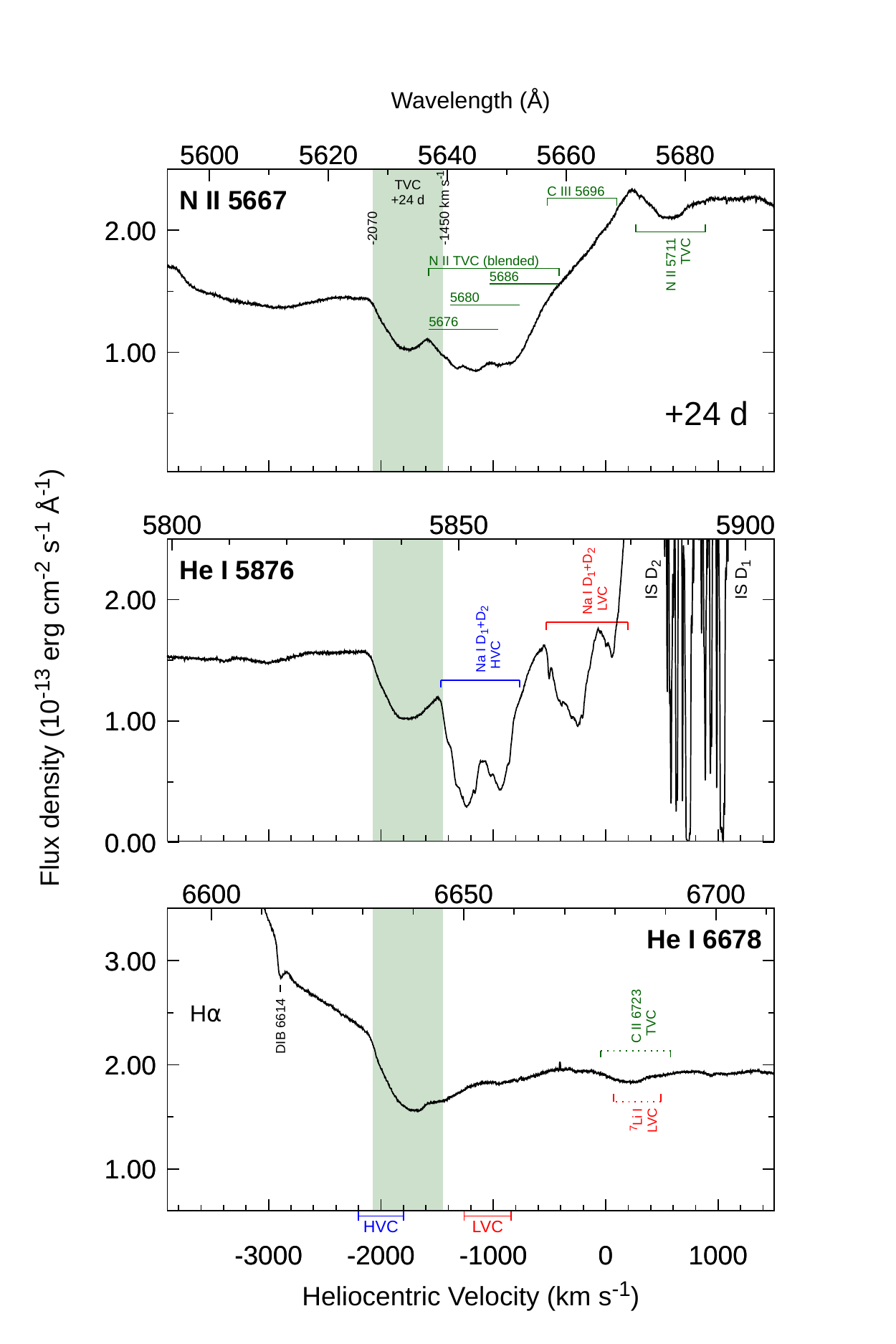}{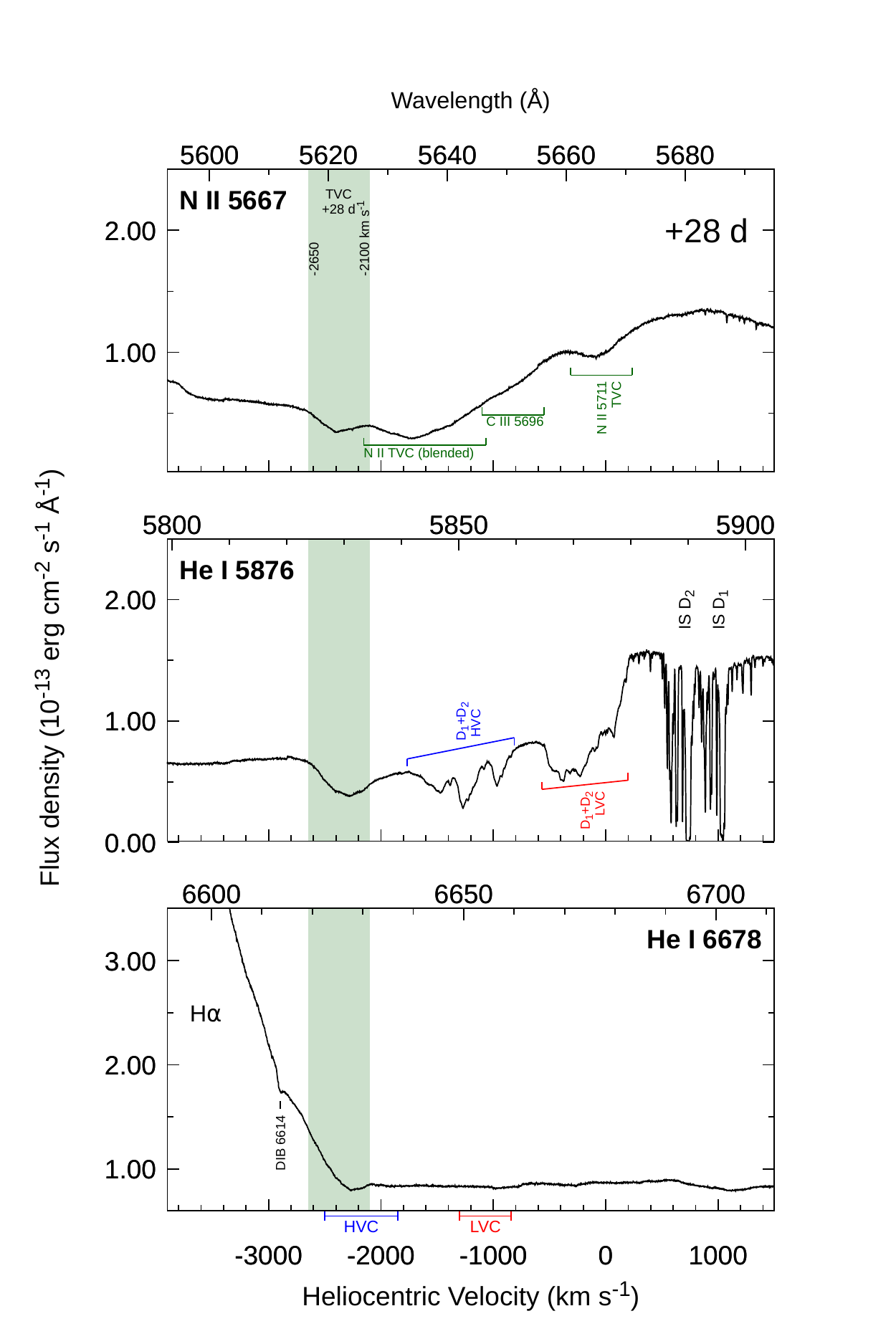}
\caption{
Temporal developments of the absorption component in the TVC (\ion{N}{2} $\lambda5666.63$, \ion{He}{1} $\lambda5875.621$, $\lambda6678.151$).
The left and right panels correspond to those on $+24$ and $+28$\,d, respectively.  
The green shaded regions denote the velocity ranges of the TVC (Section~\ref{sect4.3}) for these lines.  
Green horizontal lines with labels indicate expected positions of the TVC originating from other lines. 
Red and blue ones are LVC and HVC for other lines, respectively.
The horizontal red and blue lines below the bottom axis show the velocity ranges of LVC and HVC on both epochs, respectively.
\label{fig4}}
\end{figure}

As shown in Figure~\ref{fig3}, we can identify absorption lines belonging to the LVC and HVC systems of \BeII\ for the spectra on both $+24$ and $+28$\,d.   
Following the previous studies of \BeII\ \citep{tajitsu15,tajitsu16}, the \Be\ abundance in the LVC and HVC can be estimated by the ratios of $W$s between \BeII\ $\lambda\lambda3131$ and \ion{Ca}{2}\,K lines.

Prior to our abundance estimations, we evaluated contaminations from other metal lines to the $W$s of \BeII\ $\lambda\lambda3131$. 
No remarkable contamination was expected for \ion{Ca}{2}\,K, because no strong lines originated from Fe-peak ions around it. 
The complicated profiles of the LVC and HVC of \BeII\ indicated that \BeII\ is contaminated with some absorption lines originating from \ion{Cr}{2}\,(5), \ion{Fe}{2}\,(82), and \ion{Ti}{2}\,(67), as mentioned in the previous studies \citep{tajitsu16,molaro16, molaro2020}. 
In our spectra of \object{V5669 Sgr}, the HVC of Fe-peak elements (\ion{Fe}{2}, \ion{Ti}{2}, \ion{Cr}{2} and \ion{Mn}{2}) are generally very faint or unidentified except for strong \ion{Fe}{2} lines (e.g., \ion{Fe}{2} $\lambda5018.436$ and  $\lambda5169.028$). 
We consider that the LVCs of Fe-peak elements are the dominant source of contamination for \BeII\ lines. 
In the vicinity of \BeII\ on +24\,d, distinguishing each contaminating line is difficult due to the heavy blending of \BeII\ $\lambda3131$ and lines of Fe-peak elements. 
As we reported in Section~\ref{sect4.2}, the LVC and HVC of Fe-peak elements (\ion{Fe}{2}, \ion{Ti}{2}, \ion{Cr}{2} and \ion{Mn}{2}) on $+28$\,d are weaker than those on $+24$\,d. 
Therefore, hereafter, we estimate the \Be\ abundance using the spectrum on $+28$\,d. 
In the spectrum on $+28$\,d, we can identify the unblended LVC of \ion{Cr}{2}\,(5) $\lambda3124.978$ $[E_{\rm low} = 2.45$ eV, $E_{\rm per} = 6.42$ eV, log$(gf)=0.286$], whose $W$ is $0.55\pm0.02$\,\AA, which is between the LVC and HVC of \BeII\ (Figure~\ref{fig3}). 
Other candidates of the contaminating lines originating from Fe-peak elements in this region are \ion{Fe}{2} (82) ($E_{\rm low}=3.9$ eV, $E_{\rm up}=7.8$ eV) at $3135.36$\,\AA\ [log$(gf) = -1.13$], $3144.75$\,\AA\  $(-1.74)$ as shown in Table~2 in \cite{molaro16}, and \ion{Ti}{2}\,(67) ($E_{\rm low}=1.2$ eV, $E_{\rm up}=5.2$ eV), $3106.25$\,\AA\ $(-0.17)$, $3117.68$\,\AA\ $(-0.50)$, $3119.83$\,\AA\ $(-0.46)$ as shown in Table~2 in \cite{tajitsu16}.
Their log$(gf)$ values are smaller than those of \ion{Cr}{2}\,(5) [log$(gf) = -0.102$ -- $+0.437$].
Furthermore, we identify no significant lines originated from \ion{Fe}{2}\,(82) and \ion{Ti}{2}\,(67) on $+28$\,d even in the wavelength region free from the LVC and HVC of \BeII.
Therefore, we ignored contaminations from these multiplet lines.
We estimated $W$s of nearby \ion{Cr}{2}\,(5) lines using their log($gf$) values and the measured $W$ of \ion{Cr}{2}\,(5) $\lambda3124.978$. 
The estimated $W$s of other \ion{Cr}{2}\,(5) lines are summarized in Table~\ref{tab:contami} with the absorption components of \BeII\ (LVC or HVC) that they may contaminate. 
The total $W$s of \ion{Cr}{2}\,(5) contaminating \BeII\ is $0.87\pm0.02$\,\AA\ for the LVC and $0.59\pm0.03$\,\AA\ for the HVC.

The measured $W$s of \BeII\ in the spectrum on $+28$\,d are $1.58\pm0.02$\,\AA\ for the LVC and $1.22\pm0.02$\,\AA\ for the HVC.  
After excluding the contamination from \ion{Cr}{2}\,(5) lines, we obtained the intrinsic $W$s of \BeII\ as $0.76\pm0.06$\,\AA\ for LVC and $0.63\pm0.07$\,\AA\ for HVC.  
The measured $W$s of \ion{Ca}{2} K, which have no contaminations, are $1.63\pm0.03$\,\AA\ for the LVC and $0.73\pm0.03$\,\AA\ for the HVC.
Using these $W$s of \BeII\ and \CaK\ in Equation (1) in \cite{tajitsu16}, the number density ratios of $N(^7{\rm Be})/N({\rm Ca})$, was calculated.
We relied on the same three assumptions as in \citet{tajitsu15} and other previous studies. 
Those are (1) the difference in the color of the background light at $3131$\,\AA\ and $3930$\,\AA\ can be ignored, (2) all \Be\ and Ca in the nova ejecta are in their singly ionized states, and (3) the \NCaH\ in the absorbing gas cloud is the solar value.
Consequently, we obtained the total number density ratio, \NBeCa\ = $1.01\pm0.09$, and $1.87\pm0.21$ for the LVC and HVC, respectively. 
If we assume that \NCaH\ in the absorbing gas is the same as that of the solar photosphere ($2.19\times10^{-6}$; \citealt{asplund09}), we obtain \NBeH\ $= 2.2\pm0.2\times10^{-6}$ and $4.1\pm0.5\times10^{-6}$ for the LVC and HVC, respectively.
This \NBeH\ can be converted into the mass ratio, $X(^7{\rm Be})/X({\rm H}) = 1.5\pm0.1\times10^{-5}$ and $2.9\pm0.3\times10^{-5}$ for the LVC and HVC, respectively. 

In addition to the effects from line contaminations, which we have discussed above, our three assumptions adopted in this \NBeH\ estimation may cause uncertainties in the results. 
We expect that the most significant uncertainty originates from the difference in the color of the background light in the range of \BeII\ or \CaII; this is mainly caused by the overlap of the broad emission lines with the background continuum.
By polynomial function fitting, we found that the amplitudes of the undulated continuum level are $\sim 25\%$ in the nearby region of \BeII\ ($3080$\AA -- $3140$\AA).
This directly corresponds to a $\pm 25\%$ error in the measured $W$, i.e., the estimated \NBeH. 
Regarding the second assumption, the ionization potential is a little different between \BeII\ (the first and second ionization potentials; $I_1 = 9.32$, $I_2=18.21$ eV) and \CaII\ ( $I_1=6.11$ eV and $I_2=11.87$ eV; \citealt{NIST_ASD}). 
Some singly ionized iron-peak elements (Ti to Fe) are found in the LVC absorption system in \object{V5669 Sgr} (see Section~\ref{sect4.2}). 
Their ionization potentials are between those of Be and Ca  ($I_1$=$6.8$–-$7.9$ eV, $I_2=3.6$–-$18.1$ eV).
This situation is quite similar to that in \object{V339 Del} \citep{tajitsu15}. 
Therefore, we expected that the large fractions of Be and Ca are in the singly ionized states, as also in the case of \object{V5669 Sgr} on $t=28$~d.
Considering the third assumption, \NCaH s in nova yields tend to be comparable to or even higher than the solar value up to $\sim 10$\%--$20$\% in some theoretical simulations for CO novae (e.g., \citealt{jose20}) in which the gas with solar abundance is assumed to accumulate on the surface of the WD.
It is mainly caused by the consumption of hydrogen during TNR.
If we adopt this hydrogen reduction in TNR, the \NBeH\ of \object{V5669 Sgr} can be underestimated up to $10$\%--$20$\%. 
Finally, the total error caused by our assumptions is $-25\%$--$+50\%$ for \NBeH\ in Table~\ref{tab:abundance}.

All \Be\ eventually decays to \Li\ through the reaction of $^7{\rm Be}(e^-,\nu_{\rm e})^7{\rm Li}$ with a half-life, $\tau_{1/2}\sim53$ days.
Therefore the total number of \Be\ isotopes produced in the TNR, $N(^7{\rm Be})_{\rm {TNR}}$, must be equal to the final \Li\ produced in the observed outburst, $N(^7{\rm Li})_{\rm final}$. 
$N(^7{\rm Li})_{\rm final}$ can be estimated using $N(^7{\rm Be})_{\rm TNR} = N(^7{\rm Li})_{\rm final} = 2^{{t}/{\tau_{1/2}}} \times N({^7\rm Be}, t_{\rm obs})$, where $t_{\rm obs}=+28$\,d. 
The final \Li\ abundance produced by \object{V5669 Sgr} is expected to be $N$($^7$Li)/$N$(H)$_{\rm final} = 3.17\pm0.27 \times 10^{-6}$ for the LVC and $5.88\pm0.66 \times 10^{-6}$ for the HVC. 
Here, we use $W$(\ion{Ca}{2}~K)s for the LVC and HVC as the weight for mass ratios between them, then get a weighted $N$($^7$Li)/$N$(H)$_{\rm final}$ for the nova as $4.0\pm0.7\times 10^{-6}$. 
We will use this averaged $N$($^7$Li)/$N$(H)$_{\rm final}$ in our discussion hereafter, which corresponds to the mass ratio of \XLiH$_{\rm final} = 2.8\pm0.5 \times 10^{-5}$.

Our results of \Be\ abundance estimation are summarized in Table~\ref{tab:abundance}. 
The \NLiHfinal\ observed in \object{V5669 Sgr} is smaller by a factor of $13.5$ than the mean value in previous studies of other observed novae ($5.4\times10^{-5}$; \citealt{molaro2020}).
It is still $3$--$5$ dex larger than the \Li\ meteoritic value (\NLiH\ $=1.86\times10^{-9}$), and significantly larger than the solar photospheric value ($1.12\times10^{-11}$) given in \cite{asplund09}.
It is rather comparable to the upper limit of numerical predictions for CO novae by \cite{jose98} ($4.8\times10^{-6}$) and \cite{starrfield20} ($7.0\times10^{-6}$).
We compare this result in \object{V5669 Sgr} with previous studies and discuss it in Section~\ref{discussion}.

\subsection{Absence of \ion{Li}{1}\label{sect4.5}}

\begin{figure} 
\epsscale{1.0}
\plottwo{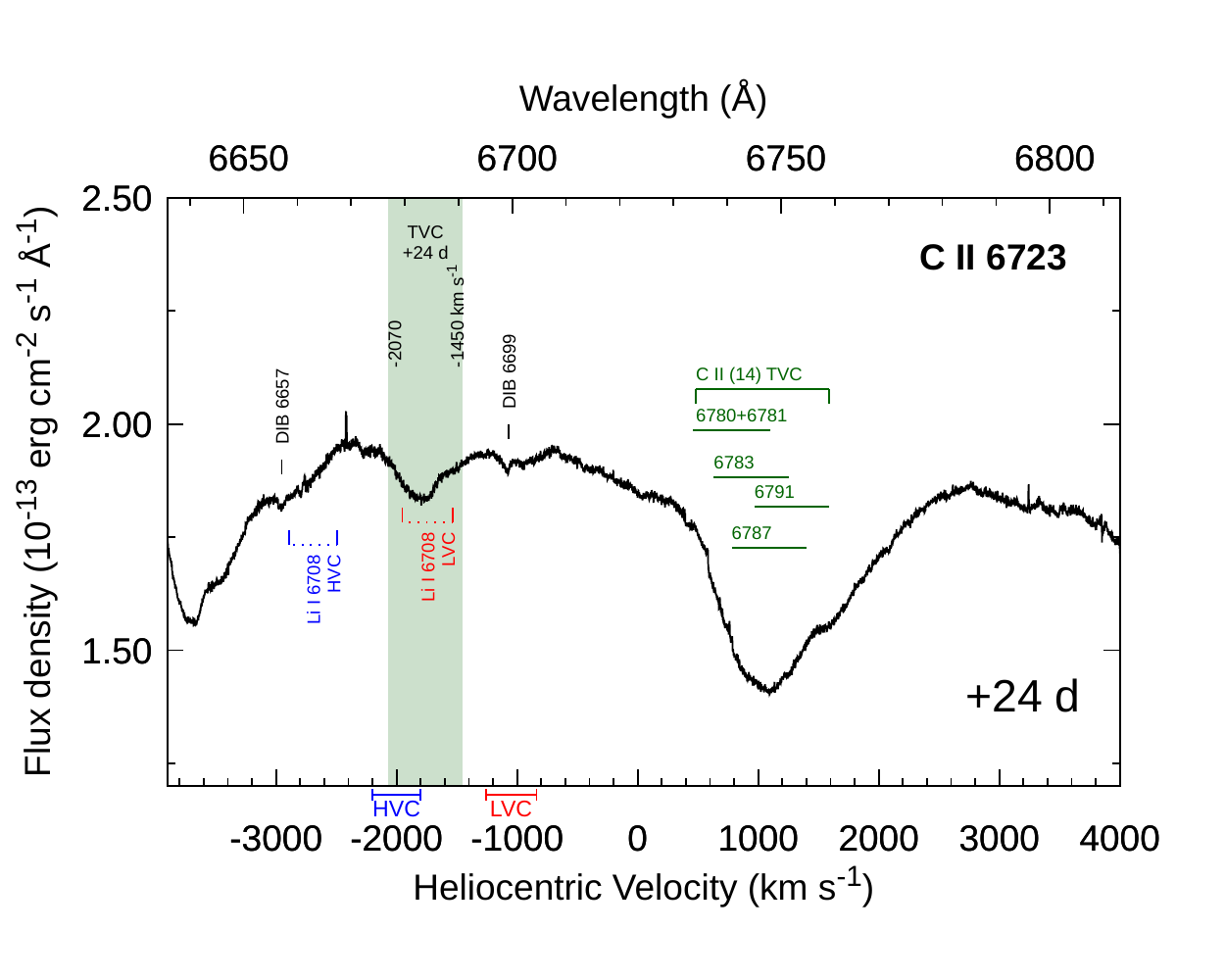}{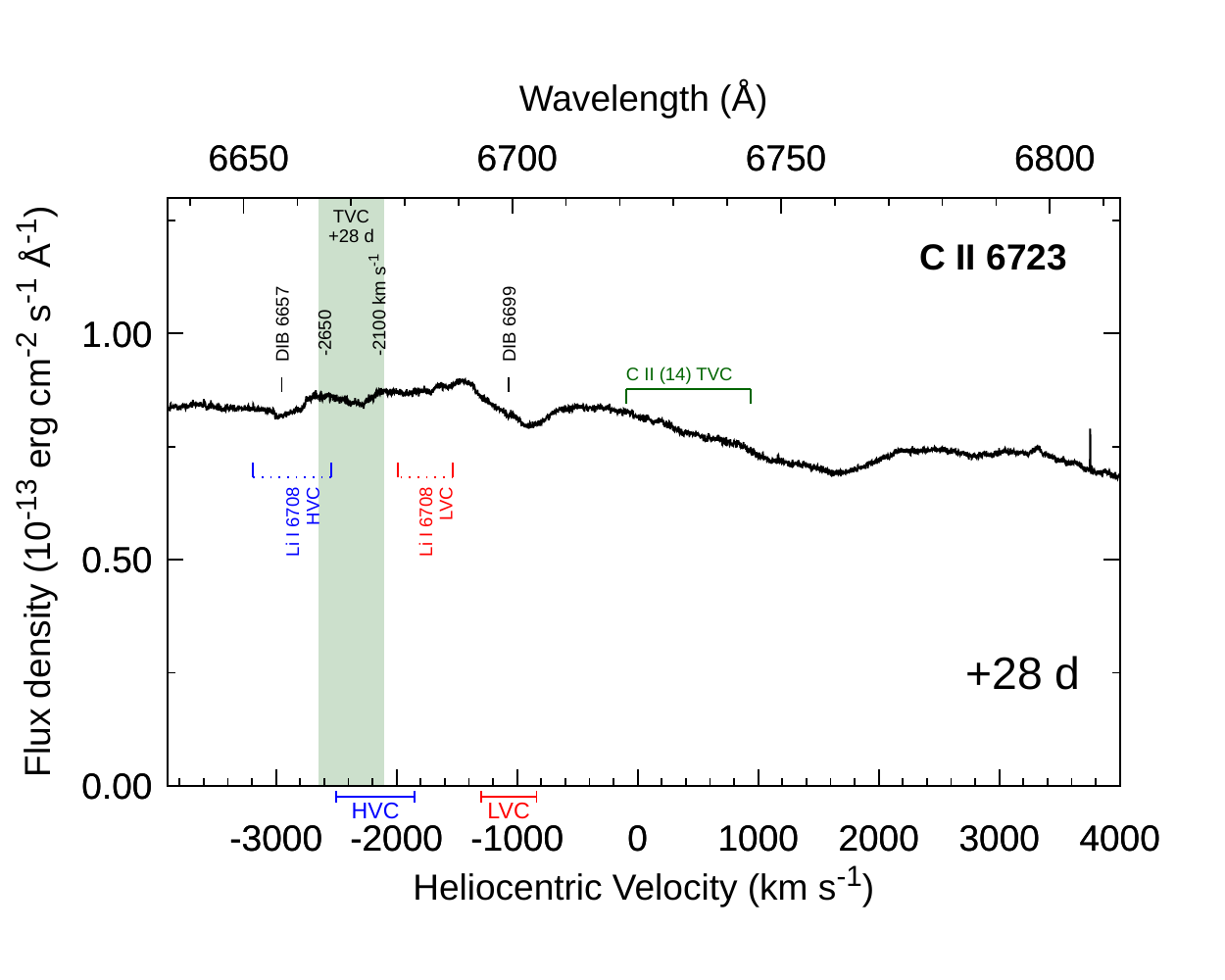}
\caption{Same as Figure~\ref{fig4}, but for the region of \ion{C}{2} $\lambda6723.32$\,\AA\ on $+24$ and $+28$\,d.   
\label{fig:CII6723}}
\end{figure}

We also checked the resonance doublet of \ion{Li}{1} at $6708$\,\AA\ although it has not been detected in most novae. 
Figure \ref{fig:CII6723} displays the spectra on $+24$ and $+28$\,d around $6683$\,\AA, where the LVC and HVC of \ion{Li}{1} $\lambda\lambda6708$ are possibly located. 
On $+24$\,d, a dip at $6683$\,\AA\ is observed.  
The position of this dip may correspond to that of LVC of \ion{Li}{1} $\lambda\lambda6708$ ($v_{\rm rad} \sim 1000$\,\kms). 
It may also agree well with the TVC of \ion{C}{2} $\lambda6723.32$ [log$(gf) = -1.156$, $E_{\rm low}=20.84$\,eV, $E_{\rm up}=22.68$\,eV].
The profile of the dip is round-shaped, which conforms to the characteristics of the TVC (Section~\ref{sect4.3}).
Furthermore, a deep and broad depression around $6750$\,\AA\ is considered to be a blending TVC arising from \ion{C}{2} (14) multiplet lines ($E_{\rm low}=20.70, E_{\rm up}=22.53$), at $6779.94$\,\AA\ [log$(gf) = +0.03$], $6780.59$\,\AA\ ($-0.38$), $6783.91$\,\AA\ ($+0.30$), $6787.21$\,\AA\ ($-0.38$), and $6791.47$\,\AA\ ($-0.27$).  
Since the $E_{\rm low}$ of \ion{C}{2} (14) and \ion{C}{2} $\lambda6723.32$ are very close, the latter's absorption strength can be estimated using that of the former by referring to their log($gf$) values. 
The measured $W$ of the blending TVC of \ion{C}{2} (14), $6.3$\,\AA, raises the expectation that the $W$ of \ion{C}{2} $\lambda6723.32$ should be $\sim 0.1$\,\AA. 
The measured $W$ of the dip at $6683$\,\AA, $\sim 0.21 \pm 0.03$\,\AA, is slightly larger than this value.
The discrepancy between them can be explained by the saturation effect of \ion{C}{2}\,(14) multiplet lines as the log($gf$) values of most multiplets in \ion{C}{2}\,(14) are much larger than that of \ion{C}{2} $\lambda6723.32$.

On $+28$\,d, there are no noteworthy dips in the range of $6679$--$6689$\,\AA, where the LVC (\vrad\ = $-840$ to $-1300$\kms) of Li I $\lambda\lambda6708$ should be located.
The TVCs (\vrad\ $= -2100$ to $-2650$ \kms) of \ion{C}{2} $\lambda6723.32$ and \ion{C}{2} (14) multiplet should locate at $\sim 6671$\,\AA\, and $6695$\,\AA, respectively.
However, both of them almost disappeared in this epoch.

\begin{figure*} 
\epsscale{1.15}
\plottwo{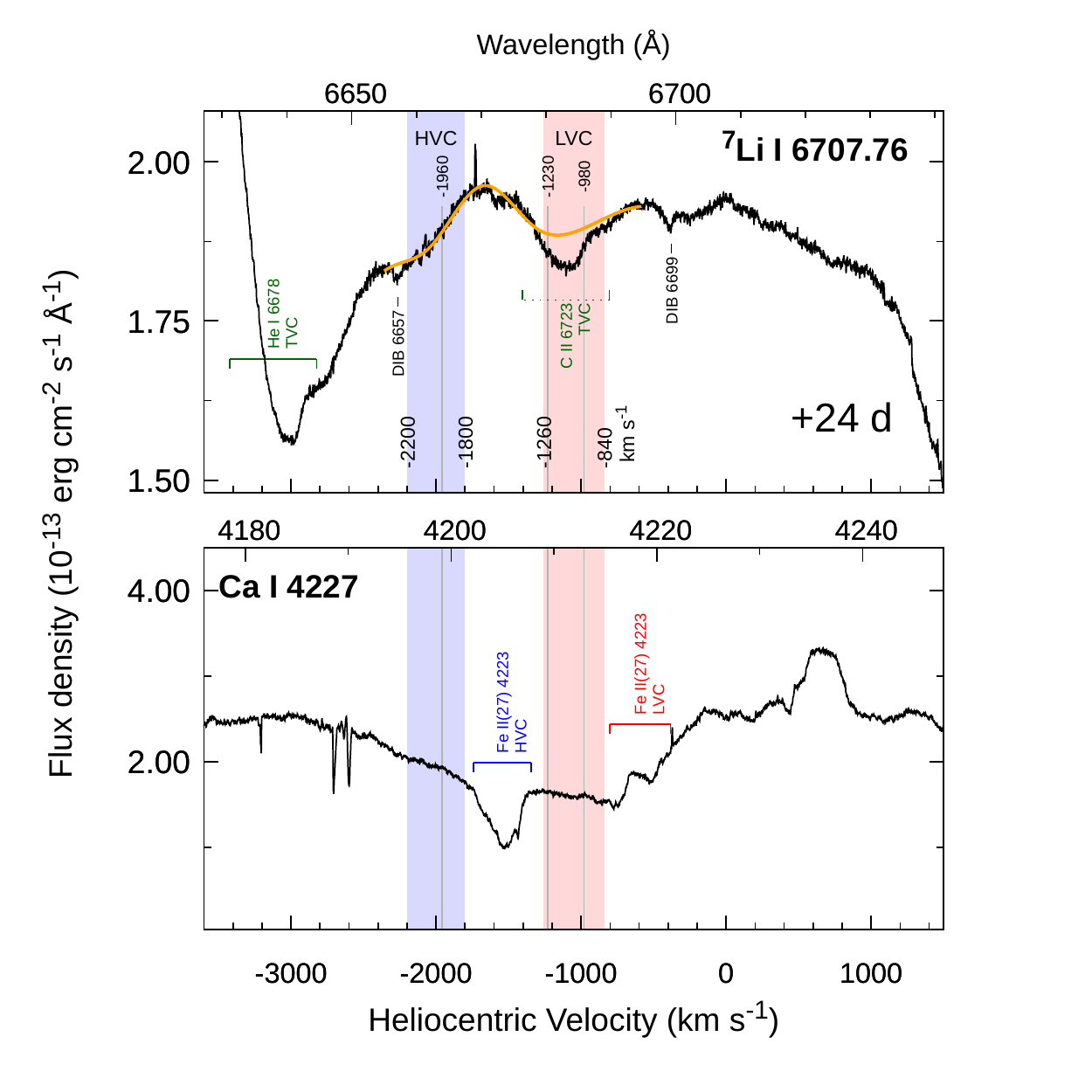}{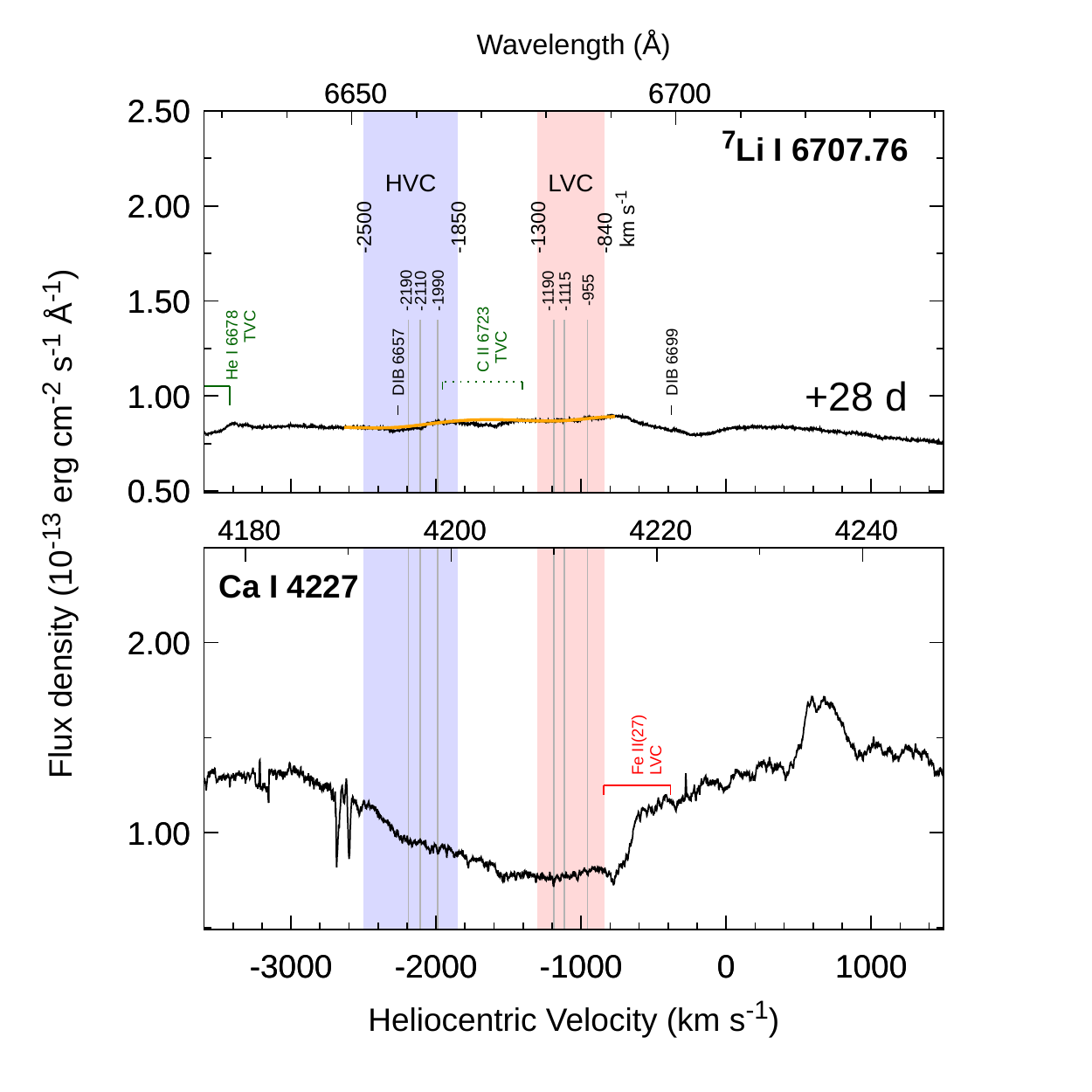}
\caption{Same as Figure~\ref{fig3}, but for regions of \ion{Ca}{1}\,$\lambda4226.73$ and \ion{Li}{1}\,$\lambda\lambda6707.76$.   No significant absorption components of \ion{Ca}{1}\, $\lambda4226.73$ are present. \label{fig:LiI-CaI}}
\end{figure*}

In the case of two novae,  \object{V1369~Cen} \citep{izzo15} and \object{ASASSN-18fv} \citep{molaro2020}, in which the \ion{Li}{1} $\lambda\lambda6708$ absorption line has been detected, the resonance absorption line of \ion{Ca}{1} $\lambda4226.73$ [log$(gf)$ = $0.244$] has been identified with \ion{Li}{1} $\lambda\lambda6708$ being in the very early stages of their outbursts (at $+7$--$+18$\,d and $1$\,d, respectively).
Figure~\ref{fig:LiI-CaI} shows the spectra of \object{V5669 Sgr} in the vicinity of \ion{Li}{1} $\lambda\lambda6708$ and \ion{Ca}{1} $\lambda4226.73$.
We found no counterparts of the LVC and HVC of \ion{Ca}{1} $\lambda4226.73$ in both epochs.
Based on these results, we conclude that the absorption of \ion{Li}{1} $\lambda\lambda6708$ is absent in our spectra of \object{V5669 Sgr} in both epochs.
We consider that the dip at $\sim 6683$\,\AA\ in the spectrum on $+24$\,d plausibly originates from \ion{C}{2} $\lambda6723.32$ instead of \ion{Li}{1} $\lambda\lambda6708$.
The absence of \ion{Li}{1} $\lambda\lambda6708$ and \ion{Ca}{1} $\lambda4226.73$ suggests that the excitation degree of ejecta observed as blue-shifted absorption components in \object{V5669 Sgr} was high enough to ionize these atoms.
This is consistent with our assumption used for our $^7$Be abundance estimation in Section~\ref{sect4.4}.

\begin{deluxetable*}{ccccccc}[]
\tablecaption{Equivalent Widths and Observed Abundances of \Be\ in \object{V5669~Sgr}.\label{tab:abundance}}
\tablecolumns{7}
\tablewidth{0pt}
\tablehead{
\colhead{} &\colhead{$W$(\BeII)} & \colhead{$W$(\ion{Ca}{2} K)} & \colhead{$\frac{N({\rm{^7Be}})}{N({\rm{Ca}})}$ } & \colhead{$\frac{N({\rm{^7Be}})}{N({\rm{H}})}$}  & \colhead{$\frac{N({\rm ^7Li})}{N({\rm H})}_{\rm final}$} & \colhead{$\frac{X({\rm ^7Li})}{X({\rm H})}_{\rm final}$}\\
\colhead{} & \colhead{(\AA)} & \colhead{(\AA)}& \colhead{} & \colhead{($10^{-6}$)} & \colhead{($10^{-6}$)} & \colhead{($10^{-5}$)}
}
\startdata
LVC   & $0.76\pm0.06$ & $1.63\pm0.03$ & $1.01\pm0.09$ & $2.2\pm0.2$ & $3.17\pm0.27$ & $2.23\pm0.19$\\
HVC   & $0.63\pm0.07$ & $0.73\pm0.03$ & $1.87\pm0.21$ & $4.1\pm0.5$ & $5.88\pm0.66$ & $4.12\pm0.46$\\
\enddata
\end{deluxetable*}


\section{Discussion\label{discussion}} 

\begin{deluxetable*}{lcccccc}[]
\tablecaption{List of \BeII\ and \LiI\ Detected Novae. \label{tab:discussion}}
\tablecolumns{7}
\tablewidth{0pt}
\tablehead{
\colhead{Object} & \colhead{$t_2$} & \colhead{$t_{\rm obs}$} & \colhead{$V_{\rm max}$} & \colhead{$M_{\rm WD}$} & \colhead{WD Type} & \colhead{\NLiHfinal} \\
\colhead{} & \colhead{} & \colhead{(d)} & \colhead{(\kms)} & \colhead{(\Msol)} & \colhead{} & \colhead{($\times 10^{-5}$)}
}
\startdata
{\bf \object{V5669 Sgr}$^{\bf (a)}$} & {\bf 24}       & {\bf 28}      & {\bf 2900}       &  {\bf 0.85}--{\bf 1.1}    & {\bf CO}   & {\bf 0.40}\\
\object{V339 Del}$^{({\rm b})}$      & 10       & 47      & 2700       &  $1.04\pm0.02^{\rm (i)}$  & CO   & $\sim2.6$\\
\object{V5668 Sgr}$^{\rm (d,e)}$     & 60       & 58, 82  & 2300       &  $0.85^{\rm (j)}$         & CO   & $\sim15$\\ 
\object{V407 Lup}$^{\rm (f)}$        & 5        & 8       & 3830       &  $0.85$--$1.35$           & ONe  & $6.2$   \\
\object{V2944 Oph}$^{\rm (d)}$       & --       & 80      & 2000       &  $0.85^{\rm (k)}$         & CO   & $1.6$   \\
\object{V838 Her}$^{\rm (g)}$        & 2        & 1, 3, 4 & 3500       &  $1.35^{\rm (j)}$         & ONe  & $2.5$   \\
\object{Nova Mus 2018}$^{\rm (h)}$   & $\sim30$ & 35      & 2240       &  $0.6$--$1.1$             & CO?  & $1.5$   \\
\object{ASASSN-18fv}$^{\rm (h)}$     & $\sim45$ & 80      & 880        &  $0.6$--$1.1$             & CO?  & $2.2$   \\
{\it \object{V1369 Cen}}$^{\rm (c,l)}$       & {\it 40} & {\it 7} & {\it 2500} &  {\it 0.9}$^{\rm (j)}$  & {\it CO} & {\it 0.015}\\
\enddata
\tablecomments{
Values of \NLiHfinal\ for all previous reported nova are quoted from \cite{molaro2020}, and averaged values are displayed regarding those of \object{V339~Del} and \object{V5668~Sgr}.\\
References: \\
(a) This study.\\
(b) \cite{tajitsu15}.\\
(c) \cite{izzo15}.\\
(d) \cite{tajitsu16}.\\
(e) \cite{molaro16}.\\
(f) \cite{izzo18}.\\
(g) \cite{selvelli18}.\\
(h) \cite{molaro2020}.\\
(i) \cite{chochol15}.\\
(j) \cite{hachisu19}.\\
(k) \cite{kato09}.\\
(l) We exclude the Li abundance of \object{V1369~Cen} in our discussion because it may not reflect the \NLiHfinal\ synthesized in this nova event, as mentioned in Section~\ref{discussion}. 
}
\end{deluxetable*}

\object{V5669 Sgr} is the eighth classical nova whose \Be\ abundance has been measured.
In addition to \object{V1369~Cen} whose $^7$Li abundance was directly measured, we summarized $^7$Li abundances of nine novae with their physical parameters in Table~\ref{tab:discussion}.
Considering the decline speed of light curves and WD masses in classical novae, we can roughly guess the WD mass of \object{V5669 Sgr} by comparing the $t_2$ of \Be\ detected novae and their WD masses reported in previous studies. %
We consider that the WD mass of \object{V5669 Sgr} ($t_{2} = 24$\,d) is larger than $0.85$ \Msol\ because the WD mass of \object{V5668 Sgr}  ($t_{2} = 40$\,d) is estimated as $0.85$ \Msol\ \citep{hachisu19}. 
We also consider that the upper limit WD of \object{V5669 Sgr} is $1.1$ \Msol, which is the highest mass of ordinary CO WD \citep{althaus10}. 
WD masses, $t_2$, and $V_{\rm max}$ of other novae are quoted from literature, as shown in the caption of Table~\ref{tab:discussion}. 
WD masses of \object{407 Lup}, \object{Nova Mus 2018} and \object{ASASSN-18fv} have not yet determined. 
However, we can roughly estimate ranges of WD mass based on their light-curve characteristics, regardless of whether they are slow or fast novae. 
WD masses for three novae are also present in Table~\ref{tab:discussion}, $0.85$--$1.35$\,\Msol\ for \object{407 Lup}, and $0.6$--$1.1$\,\Msol\ for \object{Nova Mus 2018} and \object{ASASSN-18fv}.
The measured \NLiHfinal s produced by each nova event, which should be equal to \NBeH $_{\rm TNR}$ values, are in the range of $0.45$--$15 \times10^{-5}$.
The maximum \NLiHfinal\ is observed for \object{V5668~Sgr}, and the lowest one is observed for \object{V5669 Sgr}. 
The \NLiHfinal\ of \object{V1369 Cen}, which is also the lowest on the list, is measured by the comparison of \ion{Li}{1} $\lambda\lambda6708$ with \ion{Na}{1}\,D and \ion{K}{1} $\lambda7699$ absorption lines, as in \cite{izzo15}. 
The ionization state of ejecta in \object{V1369~Cen} when \ion{Li}{1} is detected ($t = 7$\,d) should be lower than that of \BeII\ detected novae as described in Section~\ref{sect4.4}. 
However, Li (the first ionization potential, $5.39$ eV) should be partially ionized in the nova ejecta because Fe ($7.87$ eV) and Cr ($6.77$ eV) have been observed in the same absorption component. 
Due to a lack of \BeII\ observations, the measured Li abundance may not reflect the \NLiHfinal\ synthesized in this nova event. 
Therefore, we exclude the Li abundance of \object{V1369~Cen} in the following discussion.

Here, we compare the observed \NLiHfinal\ with those of the theoretical simulations.
Because WD masses are one of the key parameters in TNR simulations, resultant yields are often demonstrated as their functions. 
In Figure 7, the \NLiHfinal s observed in the nine novae are plotted 
with their estimated WD mass over the results from several theoretical simulations (\citealt{jose98,jose20,starrfield20}).
Though the errors of estimated WD masses for the observed novae in Table~\ref{tab:discussion} may be quite huge, it is remarkable that all the observed  \NLiHfinal s are higher than the results of the theoretical simulations in Figure 7.
As we discussed, the error in our \NLiHfinal\ in \object{V5669 Sgr} is $\sim \pm 0.1 \times 10^{-5}$.
The scattering of all the observed \NLiHfinal s is considerably large ($\sim15\times 10^{-5}$) compared to such errors. 
This means that the \Li\ productivity must vary in each nova. 
The \NLiHfinal\ of \object{V5669 Sgr} is the lowest among the observed novae and comparable to the upper limit of the theoretical estimations with the WD mass in the range of  $0.85$--$1.1$\,\Msol\ (\citealt{jose98,starrfield20}). 
The discrepancy in the \Li\ yields between the observations and theoretical simulations should still be discussed in future investigations.
Furthermore, the \Li\ productivity varies greatly among each existing theoretical simulation.
These imply that some additional physical mechanisms need to be considered in theoretical
simulations.
Moreover, further observational studies are required to understand the variance of Li production in classical novae. 
\cite{molaro2020} reported about two novae without \Li\ production. 
Therefore, pursuing the appearance rate of \Li\ production is quite important as well.

In this paper, we estimated the $^7$Li abundance of \object{V5669 Sgr}, which is the lowest value among other \Be\ detected novae. 
Moreover, the value agrees with theoretical predictions of the WD mass range of $0.85$--$1.1$\,\Msol, which is an order of magnitude less than those of previous observations. 
Our result indicates that there is a scattering in \Li\ production among classical novae as shown in \cite{molaro2020} who  reported a nondetection of \BeII\ lines and very low abundances of \Li\ in two novae. 
Further observational studies for the \Li\ abundances in classical novae are required to resolve the scattering of \Li\ production among them. 
This must be key to more precisely understanding the Galactic Li evolution.

\vspace{1cm}
\begin{acknowledgments}
This study is based on data collected at the Subaru Telescope, which is operated by the National Astronomical Observatory of Japan. We are honored and grateful for the opportunity of observing the Universe from Maunakea, which has the cultural, historical and natural significance in Hawaii. We gratefully thank the collaborators of our proposal for observations by the Subaru telescope, Dr. Kozo Sadakane (Osaka Kyoiku University), Dr. Hiroyuki Naito (Nayoro Observatory), Dr. Wako Aoki (National Astronomical Observatory of Japan), Dr. Satoshi Honda (Nishi-Harima Astronomical Observatory, University of Hyogo), and Mr. Mitsugu Fujii (Fujii-Kurosaki Observatory). We are also grateful for the variable star observations from the AAVSO International Database contributed by observers worldwide and that were used in this study.  This study is supported by KAKENHI (Grant-in-Aid for Challenging Exploratory Research, 15K13466 and for Scientific Research (C) 19K03933) from the Japan Society for the Promotion of Science (JSPS) and the Supported Programs for the Strategic Research Foundation at Private Universities (No. S1411028) from the Ministry of Education, Culture, Sports, Science and Technology (MEXT) of Japan.
\end{acknowledgments}

\appendix
\section{Estimations of $E_{B-V}$ and Distance of V5669 Sgr}\label{appendix}

We estimated $E_{B-V}$ of \object{V5669 Sgr} using the following three methods: 
(1) the intrinsic color index at the maximum of classical novae given by \citet{van87}; 
(2) the relation between $W$s of the diffuse interstellar band (DIB) and the interstellar hydrogen column density as practiced for \object{V407 Lup} \citep{izzo18};  
and (3) the relation between $E_{B-V}$ and $W$s of the DIB $\lambda6613.6$ among classical novae \citep{munari14}.

The intrinsic color index at the maximum of classical novae is known to be $(B-V)_{\rm max}=0.23\pm0.06$ \citep{van87}.  Then, we obtain the intrinsic extinction $E_{B-V}$ from the observed color index at the maximum, $(B-V)_{\rm obs}=0.793\pm0.001$ (SMARTS Nova ATLAS), and hence, $E_{B-V}=0.56\pm0.06$.  

Furthermore, we estimate $E_{B-V}$ of \object{V5669 Sgr} using DIBs, adopting the same procedure used for \object{V407 Lup} in \cite{izzo18}. 
DIBs, and interstellar atomic absorption lines (IS) are also useful for estimating interstellar extinction.  
DIBs ($\lambda5705.1$, $\lambda5780.5$, $\lambda5797.1$, $\lambda6196.0$, and $\lambda6613.6$) and IS (\ion{Na}{1} D$_1$ D$_2$, \ion{Ca}{2} H K, \ion{Ti}{2} $\lambda3242$, and \ion{Ti}{2} $\lambda3383$) are detected in our data on $+24$ and $+28$\,d,  because major components of IS of \ion{Na}{1} D$_1$ D$_2$ and \ion{Ca}{2} H K are heavily saturated, and DIB $\lambda5797$ and $\lambda6196$ are blended with other unidentified absorption lines in our spectra.   
We used two DIBs at $\lambda6614$ and $\lambda5780.5$ to estimate the extinction of \object{V5669}{Sgr}. Measurable DIBs and IS are listed in Table~\ref{tab:DIB}.   
The correlation coefficients between the hydrogen column density of interstellar gas ($N_{\rm H}$) and these DIBs are given in Table~2 of \cite{friedman11}.  The averaged value of $N(\rm{H})=2.06\pm0.26 \times 10^{21}\,{\rm cm}^{-2}$, affording $E_{B-V}=0.36\pm0.05$, where we adopt the general value for the interstellar hydrogen total column density of $N({\rm H})/E_{B-V}=5.87 \times 10^{21}\,{\rm cm}^{-2} {\rm mag}^{-1}$ \citep{bohlin78}.  
Both $W$s of DIBs and IS \ion{Ti}{2} of \object{V5669 Sgr}, which depend on the amount of IS should be larger than those of the \object{V407 Lup} \citep{izzo15}. 
Since we can only measure two DIBs, we use $W$s of \ion{Ti}{2} $\lambda3242$ and $\lambda3383$ listed in Table~\ref{tab:DIB} for our $E_{B-V}$ estimation.
Both $Ws$ of these \ion{Ti}{2} lines and two DIBs in \object{V5669 Sgr} are larger than those in \object{V407~Lup} ($W=101.0\pm5.8$\,\AA\ and $W=152.1\pm6.2$\,\AA\ for \ion{Ti}{2} $\lambda3242$ and \ion{Ti}{2} $\lambda3383$, respectively; \citealt{izzo18}). 

\cite{munari14} reported the empirical relation between $E_{B-V}$ and $W$s of DIB $\lambda6614$ ($W_{\lambda6614}$), $E_{B-V}=4.40 \times W_{\lambda6614}$.  
Adopting this relationship for measuring the measured $W$ of DIB $\lambda6614$ in \object{V5669 Sgr}, we obtain $E_{B-V}=0.41 \pm 0.01$. 
The $E_{B-V}$ estimated from DIBs $\lambda6614$ agrees with their standard deviations.  

Finally, we obtained the average of extinction, $E_{B-V} = 0.44\pm0.03$, from the above three values.
The absolute magnitudes of \object{V5669 Sgr} are estimated through the the maximum-magnitudes and rate-of-decline (MMRD) relation  \citep[e.g.][]{della95,downes00,ozdonmez18}.
Then, we obtained the distance to the nova using the above $E_{B-V}$. 
We used the latest MMRD formulation for Fe II novae, suggested in \citet{ozdonmez18} with $t_2$, and obtained $M_{V}=-7.8 \pm 0.4$.  Consequently, we identified the distance, $d=10.0\pm1.9$\,kpc using the averaged $E_{B-V}$ with the assumption of $R_V=A_V/E_{B-V}=3.1$.  These results are shown in Table~\ref{tab:target}.

\begin{deluxetable}{cccc}
\tablecaption{List of Measured DIB and IS lines.\label{tab:DIB}}
\tablecolumns{4}
\tablewidth{0pt}
\tablehead{
\colhead{DIB/IS} & \colhead{$W$ (+24d)} & \colhead{$W$(+28d)} & \colhead{log$N$(H)}\\
\colhead{}       & \colhead{m\AA}       & \colhead{m\AA}      & \colhead{$\times 10^{21} {\rm cm}^{-2}$}
}
\startdata
\ion{Ti}{2} 3242 & $175.2 \pm 1.4$  & $134.8 \pm 1.2$ &    \\
\ion{Ti}{2} 3383 & $234.0 \pm 1.4$  & $227.5 \pm 1.4$ &    \\
DIB 5780.5       & $338.3 \pm 3.5$  & $338.3 \pm 3.5$ & $2.62 \pm 0.56$\\
DIB 6613.6       & $ 91.7 \pm 2.5$  & $90.4 \pm 2.6$ & $1.62 \pm 0.22$\\
\enddata
\end{deluxetable}

\clearpage
\bibliographystyle{aasjournal}
\bibliography{ref_v5669sgr.rn,references_BeLi.rn,ref_in_arai+2016.rn.bib,ref_GAIA_PanSTARRS.bib}

\begin{figure*} 
\epsscale{1.0}
\plotone{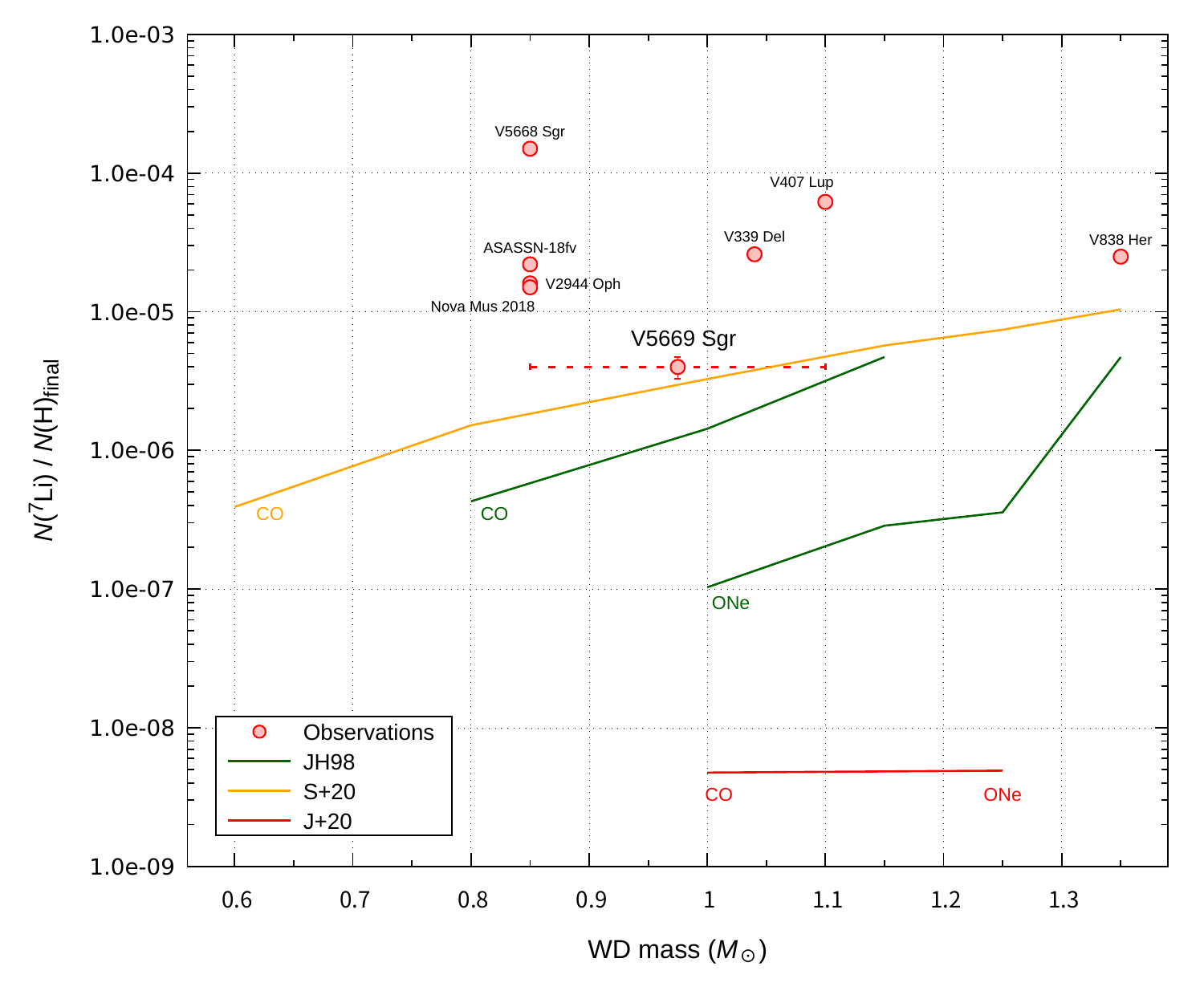}
\caption{
Comparison with observed Li abundances and recent numerical simulations along with WD masses. 
Red bordered circles denote averages of observed abundances listed in Table~\ref{tab:discussion}.
The dotted horizontal line with the value of \object{V5669 Sgr} shows the considerable WD mass range of \object{V5669 Sgr}. 
Note that WD masses for all observed novae, which have been estimated from the duration of their light curves, generally have similar large error bars.  Since evaluating errors for each WD mass estimation is difficult, we simply plotted middle points of each estimated mass ranges for each observed nova.  The green, orange, and red lines correspond to theoretical predictions; highest values for each WD mass in \citet[JH98]{jose98}, the 25\%--50\% mixing model for only CO novae of \citet[S+20]{starrfield20} and '123–-321 models' of \citet[J+20]{jose20}, respectively.
\label{fig7}}
\end{figure*}

\end{document}